# Towards quantized current arbitrary waveform synthesis

**Authors:**
P. Mirovsky, L. Fricke, F. Hohls, B. Kaestner, Ch.Leicht, K. Pierz, J. Melcher, and H.W. Schumacher

**Affiliation**
Physikalisch-Technische Bundesanstalt, Bundesallee 100, 38116 Braunschweig, Germany

Corresponding Author:     Frank Hohls
E-Mail:                   frank.hohls@ptb.de
Phone: +49 531 592 2530   Fax:    +49 531 592 69 2530

**Abstract:**

The generation of ac modulated quantized current waveforms using a semiconductor non-adiabatic single electron pump is demonstrated. In standard operation the single electron pump generates a quantized output current of $I = ef$ where $e$ is the charge of the electron and $f$ is the pumping frequency. Suitable frequency modulation of $f$ allows the generation of ac modulated output currents with different characteristics. By sinusoidal and saw tooth like modulation of $f$ accordingly modulated quantized current waveforms with kHz modulation frequencies and peak currents up to 100 pA are obtained. Such ac quantized current sources could find applications ranging from precision ac metrology to on-chip signal generation.

**Article:**

Electrical quantum metrology links the electrical units to fundamental natural constants, namely to the charge of the electron $e$ and to Planck's constant $h$ [1]. The success of quantum metrology is documented by the plans of the General Conference on Weights and Measures to redefine the international system of units (SI) based on fundamental constants [2]. In electrical quantum metrology the two effects which have been successfully applied for many years are the Josephson effect [3] and the quantum Hall effect [4]. The Josephson effect allows the generation of quantized voltages $V_J = f_J \cdot (h/2e)$ with $f_J$ being a microwave excitation frequency enabling the realization of quantum voltage standards [5]. The quantization of the Hall resistance in units of $h/e^2$, observed in a two dimensional electron gas in high magnetic fields, is used to realize quantum resistance standards [6]. Recently, semiconductor based single electron pumps (SEP) have become subject of intense studies as they might allow the realization of a quantum standard of the electrical base unit ampere [7,8,9]. When operated at a pumping frequency $f$ such single electron pumps generate quantized currents $I = ef$. These pumps have so far demonstrated output currents above 100 pA with ppm uncertainty [10], and promise lower uncertainties based on theoretical analysis [10,11,12].

Initially all electrical quantum standards have been operated only in dc mode. Later, Josephson quantized ac voltage generation was realized based on switchable binary arrays [13] and on pulse driven arrays enabling full Josephson arbitrary waveform synthesis [14,15,16,17,18] with applications ranging from ac calibrations [19] to Johnson noise thermometry [20]. Additionally studies of the ac quantum Hall effect have paved the way to a quantum standard of electrical capacitance [21].

However, the generation of quantized ac current waveforms based on frequency modulated single electron pumping has not been considered yet.

Here we carry out proof-of-principle experiments on the generation of ac modulated quantized currents using a non-adiabatic SEP [8] with frequency modulated pumping frequency. By sinusoidal and saw tooth modulation of the pumping frequency $f$ ac modulated quantized current waveforms are generated. In our experiments kHz modulation frequencies and peak currents up to 100 pA are realized. More advance modulation schemes based on pulse driven SEPs in combination with parallelization could in the future enable full arbitrary quantized current waveform synthesis with applications in metrology and on-chip signal generation.

The experimental setup is discussed with respect to Fig. 1. (a), which shows a scanning electron micrograph of a semiconductor SEP as used in the experiments. The SEP is fabricated from an about 500 nm wide wire (green) etched from an AlGaAs/GaAs heterostructure. The wire contains a high-mobility two-dimensional electron gas (2DEG) located 95 nm beneath the surface with carrier concentration $N_e =$ $2.1 \times 10^{15}$ m$^{-2}$ and electron mobility of 97 m$^2$/Vs at 4.2 K. The 2DEG channel is defined by electron-beam lithography and wet etching. Ohmic contacts to the 2DEG on the source (S) and drain (D) side are provided by alloyed AuGeNi. Three TiAu top gates $G_1$ - $G_3$ (yellow) with 100 nm width and 250 nm centre spacing cross the wire. The contacts and gates are defined by electron-beam lithography and lift off. By applying suitable negative dc voltages $V_1$-$V_3$ to $G_1$-$G_3$ the 2DEG underneath and around the gates can be depleted. While S is grounded D is connected to a low noise current-voltage converter (IVC: Femto LCA-10k-500M) generating an output voltage of 0.5 mV per pA of pumped current. The resulting output voltage waveform is sampled by an Agilent 34411A sampling voltmeter with 50 kHz sampling rate. An arbitrary waveform generator (AWG: Tektronix AWG 7102) behind a 6 db attenuator is connected to $G_1$ via a cryogenic high frequency line. This allows application of a gate voltage $V_1$ with negative dc component $V_1^{dc}$ and an additional high frequency sinusoidal ac component $V_1^{ac}$ oscillating with pumping frequency $f$. Note that $G_1$ at the end of the cryogenic high frequency line represents an open termination. Although a 50 $\Omega$ matched termination would significantly reduce signal reflections at $G_1$ it is not implemented due to the resulting high dissipated power inside the cryostat.

Negative dc voltages $V_2$, $V_3$ are applied to the centre gate $G_2$ and to the right gate $G_3$ to deplete the 2DEG underneath and in between $G_2$ and $G_3$ as sketched in Fig. 1 (b). They thus define a wide electrostatic barrier between the dynamic quantum dot (QD) between $G_1$ and $G_2$ and D. Pumping is induced by the sinusoidal voltage component $V_1^{ac}$ which induces an oscillating entry barrier between QD and S. The principle of pumping is described with respect to the three potential sketches (i)-(iii) of Fig. 1(b) which show the potential landscape during different phases of the pumping cycle. In the initial phase (i) the left barrier to source is low and highly transparent. The QD is large and is filled by electrons from source. Then (ii) the left barrier rises and becomes more opaque. At the same time the QD becomes more confined and the bound QD states are lifted above the chemical potential $\mu$ of source and drain. During this so-called decay cascade phase the QD levels of electrons with higher charging energy are still strongly tunnel coupled to source. Hence, these excess electrons tunnel back to source with tunnel rates $\Gamma >> f$ much higher than the pumping frequency $f$ resulting in a reliable initialization of the dynamic QD with one electron [11]. Finally (iii), due to further rise of the QD levels the barrier to D becomes transparent with tunnelling rate $\Gamma >> f$. Now the electron captured by the QD is ejected to D before the cycle starts over again. Under continuous AC excitation by the AWG the device generates a quantized current $I = nef$.

$n$, the number of electrons captured in the first part (i) of the pumping cycle can be controlled by adjusting the exit gate voltages $V_2$ or $V_3$. Note that in most of our previous experiments a negative dc voltage $V_2$ was applied only to $G_2$ whereas $G_3$ was grounded ($V_3$=0V) [8,12]. However, reliable single charge pumping is possible for both configurations. Also intermediate regimes of pumping through a coupled double quantum dot system have been explored [22].

Fig. 2 shows the pumping characteristics of the SEP used in the experiments. The measurements are carried out in a [3]He cryostat at base temperature 0.3 K. In Fig. 2(a) two measurements of the current $I$ generated by the SEP as function of the exit gate voltage $V_3$ in the range of -155 to -20 mV is shown. The entry gate voltage is modulated with a frequency of $f$ = 350 MHz (black) and $f$ = 100 MHz (red dashed line), respectively. The ac peak amplitude is $V_1^{ac}$ = 100 mV, $V_1^{dc}$ = -105 mV, and $V_2$ = -85 mV. The $I$ = 1$ef$ current plateaus clearly indicates reliable single electron pumping. From the shape of the plateau at $f$ = 350 MHz a theoretical current quantization accuracy of this device of the order of 500 ppm is estimated [11]. For $f$ = 100 MHz the plateau width is significantly enhanced, indicating a much lower uncertainty. Note that the best dc precision current measurement of our devices yielded uncertainties around 10 ppm limited by the measurement setup [23] while theoretical analysis of the data promises lower uncertainties well beyond 1 ppm [11,12].

Fig. 2(b) compiles measurements of the pumped current $I$ for the same gate voltage parameters for varying $f$ between 100 MHz and 600 MHz. The absolute value of the derivative of each $I(V_3)$ scan $|dI/dV_3|$ normalized by $ef$ is gray scale encoded. The light grey horizontal line at 350 MHz corresponds to the data shown in Fig. 2(a). On the left hand side of the plot ($V_3$ < -145 mV) no current is pumped and a narrow white region is found. For slightly more positive voltages of $V_3$ a gray region appears. Here pumping sets in and the current increases with $V_3$ to the $I = ef$ plateau. This plateau corresponds to the large white region in the centre part. For more positive $V_3$ the current increases beyond the plateau leading to the adjacent gray region (upper right).

For low frequencies ($f = 100$ MHz) the plateau extends over a broad region of $V_3$ from -125 to -25 mV. For higher $f$ the plateau becomes significantly narrower and for $f = 600$ MHz a quantized $I = ef$ is only generated for $V_3 = -75…-60$ mV. Note that the width of the plateau does not decrease monotonously with increasing $f$. It shows an oscillatory behaviour with a periodicity of about 42 MHz. This can be attributed to the reflection of $V_1^{ac}$ at the open termination at gate $G_1$. The reflected wave is again partially reflected at the output of the AWG leading to interference with the output waveform with a frequency dependent phase. This causes a frequency dependent variation of the effective ac amplitude at gate $G_1$ resulting in the oscillating plateau width.

Despite of the decrease of the plateau width with increasing frequency the $I(V_3) = ef$ plateau is found over the complete range of $f$ for the given gate parameters ($V_1^{dc}, V_1^{ac}, V_2$). The two vertical lines in Fig. 2(b) mark a gate voltage region of $V_3 = -72$ … -60 mV in which a quantized current $I = ef$ is generated for fixed gate voltage parameters ($V_1^{dc}, V_1^{ac}, V_2, V_3$) over the whole frequency range. Thus, in this range the generation of ac modulated quantized currents by frequency modulation of $f$ should be possible.

For these current modulation experiments, first, fixed gate voltage parameters of $V_1^{ac} = 100$ mV, $V_1^{dc} = -105$ mV, $V_2 = -85$ mV, and $V_3$ as given below are applied. Then, the frequency $f$ of the ac component of $V_1^{ac}$ is modulated by the AWG. The output waveforms consist of $10^{10}$ samples per second with 75 ps rise time between the samples.

The synthesized sinusoidal AWG output waveforms thus contain at least 16 samples per period for the maximum $f$ and can hence be considered as sufficiently smooth.

Fig. 3 shows three examples of ac modulated quantized current waveforms generated by ac modulation of the pumping frequency. Panel (a) shows a sinusoidal frequency modulation whereas (b) and (c) show saw tooth modulation with sharp falling and rising edge, respectively. In the three panels the generated current $I$ is plotted as function of time $t$ over two periods of the modulation frequency $f_m$. $f$ is modulated around the centre (carrier) frequency of 350 MHz by $\Delta f = \pm 250$ MHz and thus remains within the plateau range of $f = 100 \ldots 600$ MHz of Fig. 2. The three examples have different modulation frequencies of $f_m = 2352$ Hz (a), $f_m = 1700$ Hz (b), and $f_m = 326$ Hz (c). $V_3$ was slightly varied within the allowed range for the generation of the three curves: $V_3 = $ -60 mV (a); $V_3 = $ -62 mV (b); $V_3 = $ -70 mV (c).

Each panel of Fig. 3 contains three curves. The green dashed line corresponds to the nominal output current generated by the pump. The blue squares correspond to the measured current as derived from the sampled IVC output averaged over 500 waveforms to reduce noise. The deviation of the measured and the nominal current stems from the finite bandwidth of the IVC, i.e. from the current *measurement*. The frequency response can be well described by a first order RC circuit with a time constant of $\tau = 22$ µs. The influence of $\tau$ on the measurement results is different for different modulation frequencies $f_m$: for high frequencies above 1 kHz (a,b) a strong deviation is found whereas for lower $f_m = 326$ Hz the deviation is less pronounced. Taking this RC response into account the theoretically expected measurement curve for a given nominally generated current is computed. This theoretically expected measurement is plotted as the red dotted line. It agrees well with the experimental data proving the generation of ac modulated quantized current waveforms by frequency modulated single electron pumping.

In the present experiment the generation of ac modulated quantized current waveforms by frequency modulation of a sinusoidal drive signal $V_1^{ac}$ was demonstrated. This simple approach to frequency modulation is possible since the SEPs under consideration are robust over a wide frequency range [24]. However for a wider range of frequency modulation this approach will fail since the accuracy of the pumped current can significantly drop for high frequencies in the GHz range [10]. Furthermore, for low frequencies adiabatic back tunnelling of the electrons could increase the current uncertainty [8]. Note that the lowest current uncertainty of about 1 ppm of a non-adiabatic semiconductor SEP so far has been demonstrated in the constant current mode using a tailored pulse waveform excitation with a constant pulse repetition frequency $f_{rep}$ of about one GHz [10]. The use of such tailored pulse waveforms allows independent tuning the pumping frequency by varying $f_{rep}$ while optimizing the time dependence of the capturing process for highly accurate current quantization. For our application the use of tailored pulse waveforms in combination with frequency modulation of $f_{rep}$ could in the future allow generating ac modulated arbitrary quantized currents in the current range from zero to 160 pA with lowest current uncertainties of 1 ppm and beyond. The pulsed mode operation could then use similar modulation schemes as pulse driven Josephson arrays [14-19]. However, such optimized arbitrary quantized current generation would require an optimized experimental setup overcoming the problem of interference of the incoming and the multiple reflected drive pulse waveform, as discussed above.

Note that non-adiabatic semiconductor SEPs only allow the generation of unipolar currents where the current polarity is defined by the selection of the ac gate.

The generation of ac quantized current waveforms with oscillating polarity would thus require two pumps with opposite polarity being connected in parallel and being operated asynchronously. Alternatively a single pump with the option of asynchronous alternating modulation of both gates to (S) and (D) could be tested. The increase of the ac current amplitude could follow the approach of on-chip parallelization as demonstrated for dc currents [25,26,27]. In the future arbitrary waveform synthesis of quantized currents could find metrological applications for the calibration of picoampere ac current sources, ramp generators, and ammeters, among others. Furthermore it could allow the on-chip generation of quantized ac currents which could drive high resistive loads with application in hybrid integrated quantized circuits [12] or solid state quantum devices.

The authors thank Holger Marx for wafer growth and Thomas Weimann for electron beam lithography. This work has been supported by DFG and within the Joint Research Project "Quantum Ampere" (JRP SIB07) within the European Metrology Research Programme (EMRP). The EMRP is jointly funded by the EMRP participating countries within EURAMET and the European Union.

**Figures:**

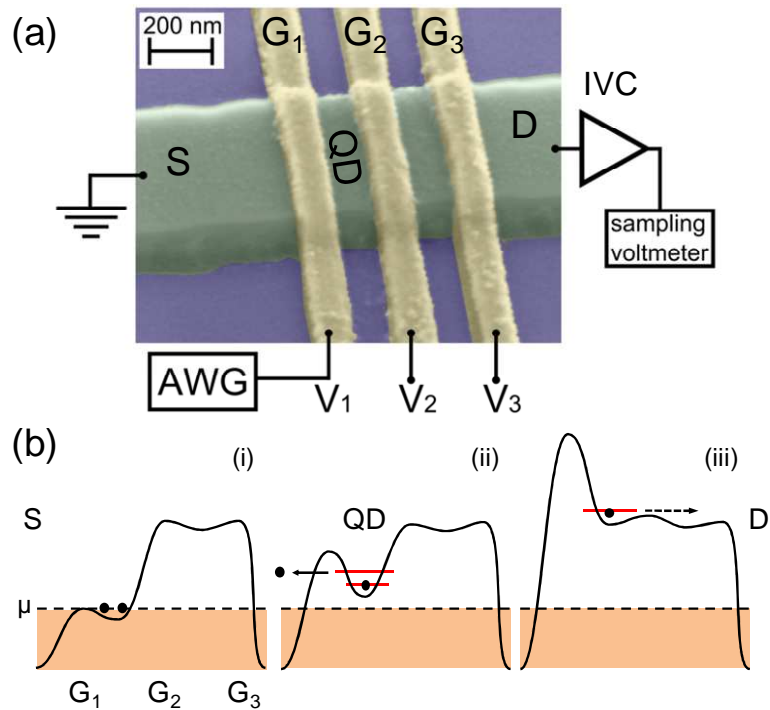

**Figure 1:**

(a) Experimental setup: Electron micrograph of SEP and electrical connections. The QD separated from source (S) and drain (D) is formed between gates $G_1$ and $G_2$. $G_1$ is connected to an arbitrary waveform generator (AWG). The current is measured by a current-voltage converter (IVC) and a sampling voltmeter. (b) Principle of pumping: The oscillating entry barrier to S defined by $G_1$ induces capturing of electrons by the QD and ejection to D.

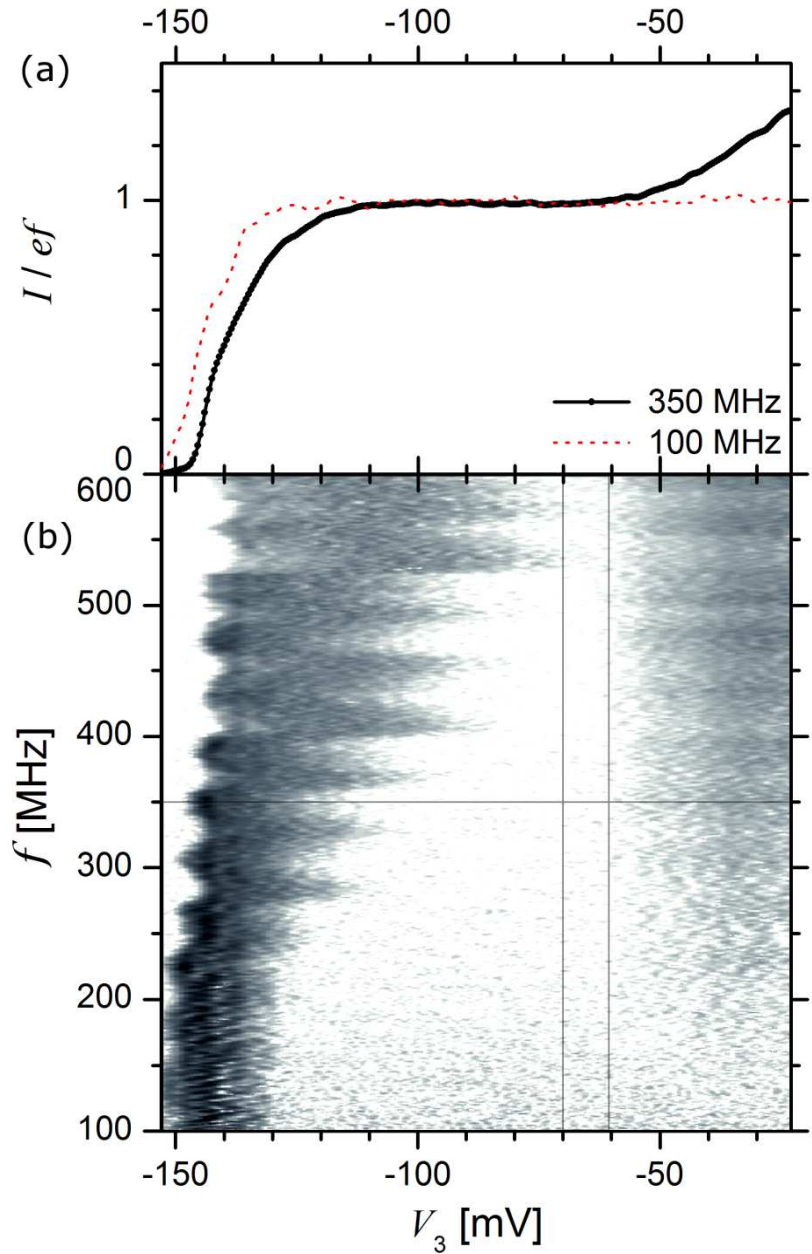

**Figure 2:**

Frequency dependent single electron pumping. (a) *I vs. $V_3$* for $f = 350$ MHz (black line) and 100 MHz (red dashed line). $V_1^{ac} = 100$ mV, $V_1^{dc} = -105$ mV, and $V_2 = -85$ mV. (b) Frequency variation of $(ef)^{-1} \cdot |dI/dV_3|$ *vs.* $V_3$ and $f$ in gray scale. White: $|dI/dV_3| \leq 5 \cdot 10^{-3}$ $ef$/mV, black: $|dI/dV_3| \geq 0.1$ $ef$/mV. White centre region: $I = ef$ plateau. For $V_3$ between the vertical gray lines $I = ef$ for complete frequency range.

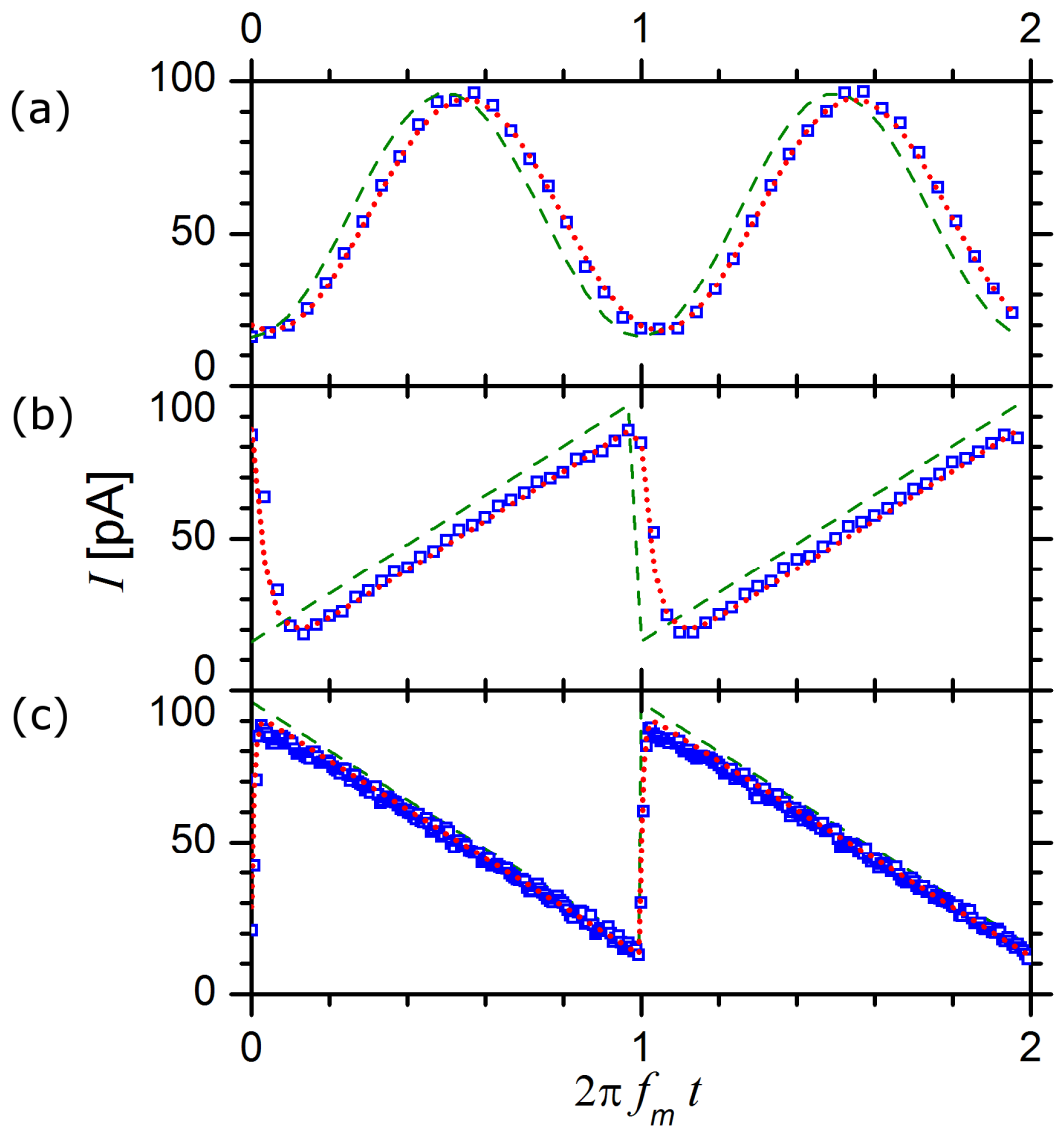

**Figure 3:**

Generation of ac modulated quantized current waveforms. (a) Sinusoidal modulation. $f_m$ = 2352 Hz, $V_3$= -60 mV. (b) Saw tooth modulation. $f_m$ = 1700 Hz, $V_3$= -62 mV. (c) Inverted saw tooth modulation. $f_m$ = 326 Hz, $V_3$ = -70 mV. Green dashed line: nominal output current. Blue squares: measured current. Red dotted line: theoretically expected measurement signal accounting for measurement bandwidth. Good agreement of the theoretically expected data and the measurement is obtained.

**References:**

[1]    I.M. Mills, P.J. Mohr, T.J. Quinn, B.N. Taylor, and E.R. Williams, Metrologia **43**, 227 (2006).

[2]    http://www.bipm.org/en/si/new_si/

[3]    B. D. Josephson, Phys. Lett. **1**, 251 (1962).

[4]    K. v. Klitzing, G. Dorda, M. Pepper, Phys. Rev. Lett. **45**, 494 (1980).

[5]    C. Hamilton et al. IEEE Electron. Device Lett. **6**, 623 (1985).

[6]    F. Delahaye et al. Metrologia **34**, 211 (1997).

[7]    M.D. Blumenthal et al., Nat. Phys. **3**, 343 (2007).

[8]    B. Kaestner et al. Phys. Rev. B **77**, 153301 (2008).

[9]    A. Fujiwara, K. Nishiguchi, Y. Ono, Appl. Phys. Lett. **92**, 042102 (2008).

[10]   S. P. Giblin et al., Nature Commun. 3, 930 (2012).

[11]   V. Kashcheyevs, B. Kaestner, Phys. Rev. Lett. **104**, 186805 (2010).

[12]   F. Hohls, et al. Phys. Rev. Lett. **109**, 056802 (2012), and supplemental material.

[13]   C. A. Hamilton, C. J. Burroughs, and R. L. Kautz, IEEE Trans. Instrum. Meas. **44**, 223 (1995).

[14]   S. P. Benz and C. A. Hamilton,  Appl. Phys. Lett. **68**, 3171 (1996).

[15]   S. P. Benz, C. J. Burroughs, and P. D. Dresselhaus, Appl. Phys. Lett. **77**, 1016 (2000).

[16]   Helko E van den Brom et al. Supercond. Sci. Technol. **20**, 413 (2007).

[17]   C. Urano et al. Supercond. Sci. Technol. **22**, 114012 (2009).

[18]   Mun-Seog Kim, et al. Meas. Sci. Technol. **21**, 115102 (2010).

[19]   R. Behr, L. Palafox, G. Ramm, H. Moser, and J. Melcher, IEEE Trans. Instrum. Meas. **56** 235 (2007).

[20]   S. P. Benz, J.M. Martinis, P.D. Dresselhaus, Sae Woo Nam, IEEE Trans. Instrum. Meas. **52**, 545 (2003).

[21]   J. Schurr, F.-J. Ahlers, B. Kibble, Meas. Sci. Technol. **23**, 124009 (2012).

[22]   P. Mirovsky et al. *Conference on Precision Electromagnetic Measurements 13,* 18 June 2010, Daejon, Korea,  10.1109/CPEM.2010.5544459.

[23]   B. Kaestner et al. *Conference on Precision Electromagnetic Measurements 14*, July 2012, Washington DC, USA.

[24]   B. Kaestner et al. Appl. Phys. Lett. **92**, 192106 (2008).

[25]   P. Mirovsky et al. Appl. Phys. Lett.  **97**, 252104 (2010).

[26]   S. J. Wright et al. Phys. Rev. B **80**, 113303 (2009).

[27]   V. F. Maisi et al. New J. Phys. **11**, 113057 (2009).